\documentclass{sig-alternate}
\usepackage[usenames,dvipsnames]{xcolor}
\usepackage{hyperref}
\usepackage{times}
\usepackage{xspace}
\usepackage{subfigure}
\usepackage[vlined, ruled]{algorithm2e}
\usepackage{multirow}
\usepackage{verbatim}
\usepackage{wrapfig}
\usepackage{booktabs}
\usepackage{balance}

\newcommand{\xourname}{Facets}
\newcommand{\name}{\textsc{\xourname}\xspace}
\newcommand{\ourmethod}{\name}
\newcommand{\ourmethodtitle}{FACETS\xspace}

\newcommand{\todo}[1]{\textcolor{red}{[#1]}}

\newcommand{\KL}{\mathit{KL}}
\newcommand{\JS}{\mathit{JS}}

\newcommand{\Nodes}{\mathit{Nodes}}

\makeatletter
\renewcommand*{\@fnsymbol}[1]{\ensuremath{\ifcase#1\or 
\circ\or 
\bullet\or 
\mathsection\or 
\ddagger\or
\mathparagraph\or 
\|\or 
**\or 
\dagger\dagger\else
\@ctrerr
\fi}}
\makeatother

\begin{document}
\setlength{\pdfpagewidth}{8.5in}
\setlength{\pdfpageheight}{11in}

\CopyrightYear{2015}

\title{Seeing the Forest through the Trees:\\
Adaptive Local Exploration of Large Graphs}

\numberofauthors{1} 
%

%

\author{
\alignauthor
Robert Pienta\footnotemark[1] \quad Zhiyuan Lin\footnotemark[1] \quad Minsuk Kahng\footnotemark[1] \quad Jilles Vreeken\footnotemark[2] \quad Partha P.\ Talukdar\footnotemark[3]  \\
 James Abello\footnotemark[4] \quad Ganesh Parameswaran\footnotemark[5] \quad Duen Horng (Polo) Chau\footnotemark[1]\\
\affaddr {\footnotemark[1]~~Georgia Institute of Technology,\; 
\footnotemark[2]~~Max Planck Institute for Informatics and Saarland University,\\ \footnotemark[3]~~Indian Institute of Science,\; 
\footnotemark[4]~~Rutgers University,\;
\footnotemark[5]~~Yahoo! Inc.} 
\email 
\{pientars,zlin48,kahng\}@gatech.edu,\; jilles@mpi-inf.mpg.de,\; ppt@serc.iisc.in, \\ abello@dimacs.rutgers.edu,\; \{ganeshcs,polo\}@gatech.edu
}

\date{20 February 2015}

\maketitle
\begin{abstract}

Visualization is a powerful paradigm for exploratory data analysis. 
Visualizing large graphs, however, often results in a meaningless hairball.
In this paper, we propose a different approach that helps the user \emph{adaptively} explore large million-node graphs from a \emph{local} perspective. 
 For nodes that the user investigates, we propose to only show the neighbors with the most subjectively interesting neighborhoods. 
 We contribute novel ideas to measure this interestingness in terms of how surprising a neighborhood is given the background distribution, as well as how well it fits the nodes the user chose to explore. 

We introduce \ourmethod, a fast and scalable method for visually exploring  large graphs. 
By implementing our above ideas, it allows users to look into the forest through its trees.
 Empirical evaluation shows that our method works very well in practice, providing rankings of nodes that match interests of users. Moreover, as it scales linearly, \ourmethod is suited for the exploration of very large graphs.

\end{abstract}

\section*{Categories and Subject Descriptors}
H.2.8\,[\textbf{Database\,management}]:\,Database\,applications--\textit{Data\,mining}



\section{Introduction}
\label{sec:intro}

Large graphs are ubiquitous. They are natural representations for many domains, and hence we find graph structured data everywhere.
As data collection becomes increasingly simple, and many domains remain complex,  real-world graphs are growing extremely large and rich with data.
These graphs may have thousands or more of attributes and scale up to and beyond millions of nodes and billions of edges. 
It is fair to say that many graphs are in fact \emph{too big}; exploring such large graphs, where the goal of the user is to gain understanding, is a highly non-trivial task.

Visualization is perhaps the most natural approach to exploratory data analysis. Under the right visualization, finding patterns, deciding what is interesting,  what is not, and what to investigate next are all easy tasks -- in a sense the answers  `jump to us' as our brains are highly specialized for analyzing complex visual data. It is therefore no surprise that Shneiderman's mantra of ``overview, zoom \& filter, details-on-demand''~\cite{shneiderman1996eyes} has proven to be successful in many domains~\cite{keim2001visual, plaisant1996lifelines, shneiderman1996eyes}. 

\begin{figure}
\centering
\includegraphics[width=0.45\textwidth]{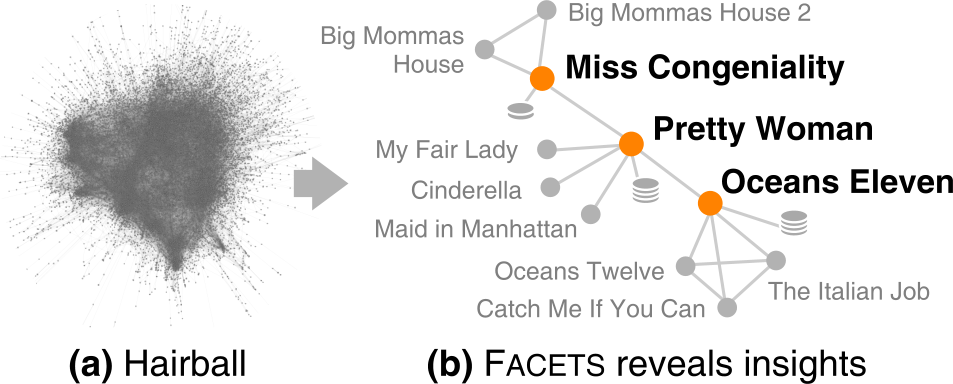}
\caption{(a) The Rotten Tomatoes movie graph shown using conventional spring layout (an edge connects two movie nodes if some users voted them as similar). Even for this relatively small graph of $17$k nodes and $72$k edges, a global visualization does not provide much insight. (b) A better way, using our \ourmethod approach, focuses on movies that are the most subjectively interesting, surprising, or both. For example,  \ourmethod suggests \textit{Pretty Woman} (romantic-comedy) as a interesting, suprising related movie of \textit{Miss Congeniality} (crime-comedy).}
\label{fig:crownjewel}
\end{figure}

Visualizing large graphs in an intuitive and informative manner has proven to be rather difficult. Even using advanced layout techniques plotting a  graph typically leads to a useless ``hairball'' from which nothing can be deduced~\cite{keim2001visual,keim2002information}. This is even the case when we plot graphs with only thousands of nodes (see Figure~\ref{fig:crownjewel}(a) for an example). Instead of plotting the whole graph, visualizing only part of the graph seems more promising~\cite{van2009search,chau2011apolo}. 
Doing this naively leads to the same problem. 
Because real world graphs are often scale free (follow a power law degree distribution~\cite{faloutsos:99:cubed}), even a single hop expansion from a selected node can be visually overwhelming.

We take a different approach. We propose to \emph{adaptively} explore large graphs from a \emph{local} perspective. That is, starting from an initially selected node -- e.g., explicitly queried by the user, or proposed by an outlier detection algorithm~\cite{akoglu2010oddball} -- we propose to only show the most interesting neighbors. We identify these by their \emph{subjective interestingness}, by how surprising their neighborhood distribution is (e.g., do neighbors' degree distributions follow a power law like when considering all nodes?), as well as by how similar this distribution is to those of the nodes the user already visited during the exploration. By only showing those parts of the graph that are most informative to the user. we keep the view clean. By being  adaptive, \ourmethod allows every user to explore those facets of the graph that are most interesting to them. 

We call our adaptive approach \ourmethod\ --- our idea is a significant addition to existing works that aim to recommend individual nodes to users; instead, we steer users towards local regions of the graphs that match best with their current browsing interests, helping them better understand and visualize the graph at the same time.

To illustrate how \ourmethod works in practice, consider our user Susan who is looking for interesting movies to watch (see Figure~\ref{fig:crownjewel}), by exploring a Rotten Tomatoes movie similarity graph with 17k movies.
In this graph, an edge connects two movie nodes if users of RottenTomatoes voted them as similar films. 
Susan has watched \textit{Miss Congeniality}\footnote{One of the most frequently rated movies on Netflix}, a crime-comedy that stars Sandra Bullock as an FBI agent who thwarts terrorist efforts by going undercover, turning her rude unflattering self into a glamorous beauty queen (see Figure~\ref{fig:crownjewel}b).
\ourmethod simultaneously suggest a few movies that are interesting and surprising to her.

Matching Susan's interest, \ourmethod suggests the \textit{Big Mommas House} series, which also have undercover plots and are interestingly like \textit{Miss Congeniality}. They both share low critics scores, but high audience scores (i.e., most critics do not like them, but people love them).
To Susan's surprise, \ourmethod suggests \textit{Pretty Woman}, which is quite different (thus surprising) --- a romantic-comedy that has both scores from the critics and the audience.
But, there is still more subtle similarity (thus still drawing Susan's interest);  both films share a Cinderalla-like storyline, which explains why the two movies are connected in the graph: Sandra Bullock goes from a rude agent to a beauty queen; in \textit{Pretty Woman}, Julia Roberts goes from a prostitute to a fair lady.
In fact,  \textit{Pretty Woman} is a classic, exemplar romantic-comedy; many movies follow similar story lines (e.g., \textit{Maid in Manhattan}). Thus, \textit{Pretty Woman} has very high degree in the graph, unlike \textit{Miss Congeniality} which is a niche genre; this also contributes to \textit{Pretty Woman}'s surprisingness.

Through \textit{Pretty Woman}, \ourmethod again pleasantly surprises Susan with \textit{Oceans Eleven}, which also stars Julia Roberts, and is in a rather different light hearted crime or heist genre, introducing Susan to other very similar movies like \textit{Oceans Twelve} and \textit{The Italian Job}. Figure \ref{fig:crownjewel}(b) summarizes Susan's exploration. If Susan were to use a conventional visualization tool to perform the same kind of movie exploration, she would likely be completely overwhelmed with an incomprehensible hairball visualization (as in Figure~\ref{fig:crownjewel}(a)).


The key contributions of our work include:
\begin{itemize}
\item A framework for locally exploring a graph without clutter, showing only the most subjectively most interesting nodes, and hence being \emph{adaptive} to the users' interests.
\item A formal notion of subjective interestingness for graph exploration taking both divergence between local and global distributions, and similarity to explored nodes into account. 
\item A measure of surprise over neighborhoods -- rather than local node attributes -- to draw users in the direction of graph areas with subjectively interesting content.
\item A highly scalable method, \ourmethod, for adaptively exploring very large graphs in a visual environment. Experimental evidence demonstrates the effectiveness of \ourmethod.
\end{itemize}

The rest of the paper is organized as follows. 
First, in Sec.~\ref{sec:model} we formalize the problem, introduces our notions of interestingness, and propose our \ourmethod solution. 
Then, in Sec.~\ref{sec:interface} we present our ideas as an integrated approach, with visualization and algorithms working closely together. 
We empirically evaluate in Sec.~\ref{sec:experiment}. 
We discuss related work in Sec.~\ref{sec:related}, and end with conclusions in Sec.~\ref{sec:conclude}.

\section{\ourmethodtitle: \\Adaptive Graph Exploration}
\label{sec:model}
In this section, we formalize the problems we aim to solve through our \ourmethod approach to achieve adaptive exploration. Then, we describe our main ideas, and describe our solutions.

To enhance readability, we have listed the symbols used in this paper in Table \ref{table:notation}. 
The reader may want to return to this table for technical terms meanings used in various contexts of discussion. 

\begin{table}[h]
\centering
\begin{tabular}{c l}
\toprule
\textbf{Symbol} & \textbf{Description}\\ 
\midrule
$v_i$ & Node $i$ \\
$D_{\JS}$ & Jensen-Shannon Divergence \\
$D_{\KL}$ & Kullback-Leibler Divergence\\
$s_i$ & surprise-score for node $v_i$\\
$r_i$ & interest-score for node $v_i$\\
$\hat S_a$ & surprise scores for all neighbors of $v_a$\\
$\hat R_a$ & interest scores for all neighbors of $v_a$\\
$f_j$ & a $j$-th feature for nodes\\
$\lambda_j$ & weight of feature $f_j$ \\
$L_{i,j}$ & neighborhood dist. of node $v_i$ for feature $f_j$\\
$G_{j}$ & global dist. for feature $f_j$\\
$U_j$ & user profile dist. for feature $f_j$ \\ 
\bottomrule
\end{tabular}
\caption{Symbols \& Notation}
\label{table:notation}
\end{table}

\subsection{Problem Definition}
The input is a graph $G=(V, E, A)$ where $A$ is a set of numerical or categorical attributes. Each node $v_i \in V$ has a corresponding attribute value for each attribute (feature) $f_j \in A$ (e.g., degree). 
Our approach works with both categorical and numerical attributes.
We assume there are no self-edges.

We solve the following problem with \name: 
\newdef{definition}{Definition}
\begin{definition}
\textbf{Node-Sequence Aware Ranking.}
Given a starting node $v_a$,  a sequence of nodes $V_h \subset V$ in which a user has shown interest,
how can we find the top-$k$ nodes among the neighbors of $v_a$ that balance
(1) similarity by features to the sequence of $V_h$ nodes (subjective interest) and (2) uncommon compared to the global distribution (surprise).  
\end{definition}

Graph exploration is an interactive and iterative process, where the user incrementally explore larger parts of the graph.  \ourmethod solves the above problem repeatedly.

A common approach to rank nodes is by their \textit{importance} scores, which are often computed using PageRank \cite{page1999pr}, Personalized PageRank \cite{haveliwala2002ppr} or random walk with restart \cite{tong2006rwr}.

However, there are other ways to rank the nodes, like using surprise or interest \cite{onuma2009tangent, akoglu2010oddball}. 
We have chosen to rank nodes by their  surprise and user-driven interest rather than by the more conventional importance metrics.
We chose surprise, because serendipitous results and insight do not always come from the most topologically important nodes \cite{onuma2009tangent}.
We made \name adaptive, because what makes a nodes interesting varies from person to person and so too should be the criteria for ranking.
For each node we suggest a combination of the most surprising and most interesting neighbors at each step of the journey.

\vfill
\subsection{Feature Distributions}
\name uses feature-based surprise and interest in order to guide the graph exploration process.
Even when a dataset does not contain node-level features, we can derive node features by using established approaches like PageRank, centrality measures or labels drawn from clustering approaches.
This means that even without a set of initial features, it's still possible for \name to guide graph exploration.

\name requires a compact representation of feature distributions.
Histograms are a natural and computationally inexpensive way to represent distributions. Our approach can consider any histogram, regardless of the binning strategy -- e.g., equi-width or equi-height binning -- used to infer the histogram. Here, we opt to use the parameter-free technique by Kontkanen and Myllymaki~\cite{kontkanen:07:histo} that is based on the Minimum Description Length (MDL) principle. In a nutshell, it identifies as the best binning the one that best balances the complexity of the histogram and the likelihood of the data under this binning. In practice this means it automatically chooses both the number of and locations for the cut points that define the histogram depending on the complexity and size of the data.

\begin{definition}
\textbf{Representing Local Feature Distributions.}
We first create a histogram for a given feature $f_j$ and a set of nodes $V$ with their feature values.
A histogram consists a set of bins $b \in B_j$ each of which has a probability value based on the number of nodes corresponding to. 
Although we have chosen MDL binning to construct our histograms, \name will work with most histograms and binning approaches.

The \textbf{neighborhood (or local) distribution} $L_{i,j}$ is a distribution of features $f_j$ over a set of neighbors of a particular node $v_i$;
The \textbf{global distribution} $G_{j}$ is the feature distribution across all nodes; and 
the \textbf{user profile distribution} $U_j$ is the distribution of a sequence of interesting nodes $V_h$ collected from the user during interaction with \name.
\end{definition}

\name works by guiding users during their graph exploration using both surprisingness (Section \ref{sec:surprise}) and subjective interest that changes dynamically to suit the user (Section \ref{sec:interest}).
We do this comparing the local or neighborhood feature distributions with the global to determine surprisingness and the local with a user profile to determine dynamic subjective interest.

\subsection{Ranking by Surprise\label{sec:surprise}}

In order to calculate a node's surprisingness we compare the distribution of the node's neighbors with the global distribution for each feature.
We chose a combined feature-centric and structural approach, because both structure and features play a critical role in inference problems \cite{lee2013attribute}.
Nodes whose local neighborhood vary greatly from the global are likely to be more surprising as they do not follow the general global trends.

One approach is to use the base entropy over node features to detect anomalous nodes; however, this ends up biasing the ranking towards skewed distribution.
Instead we measure the difference between two distributions for more consistent results.
\begin{figure}
\centering
\includegraphics[width=.45 \textwidth]{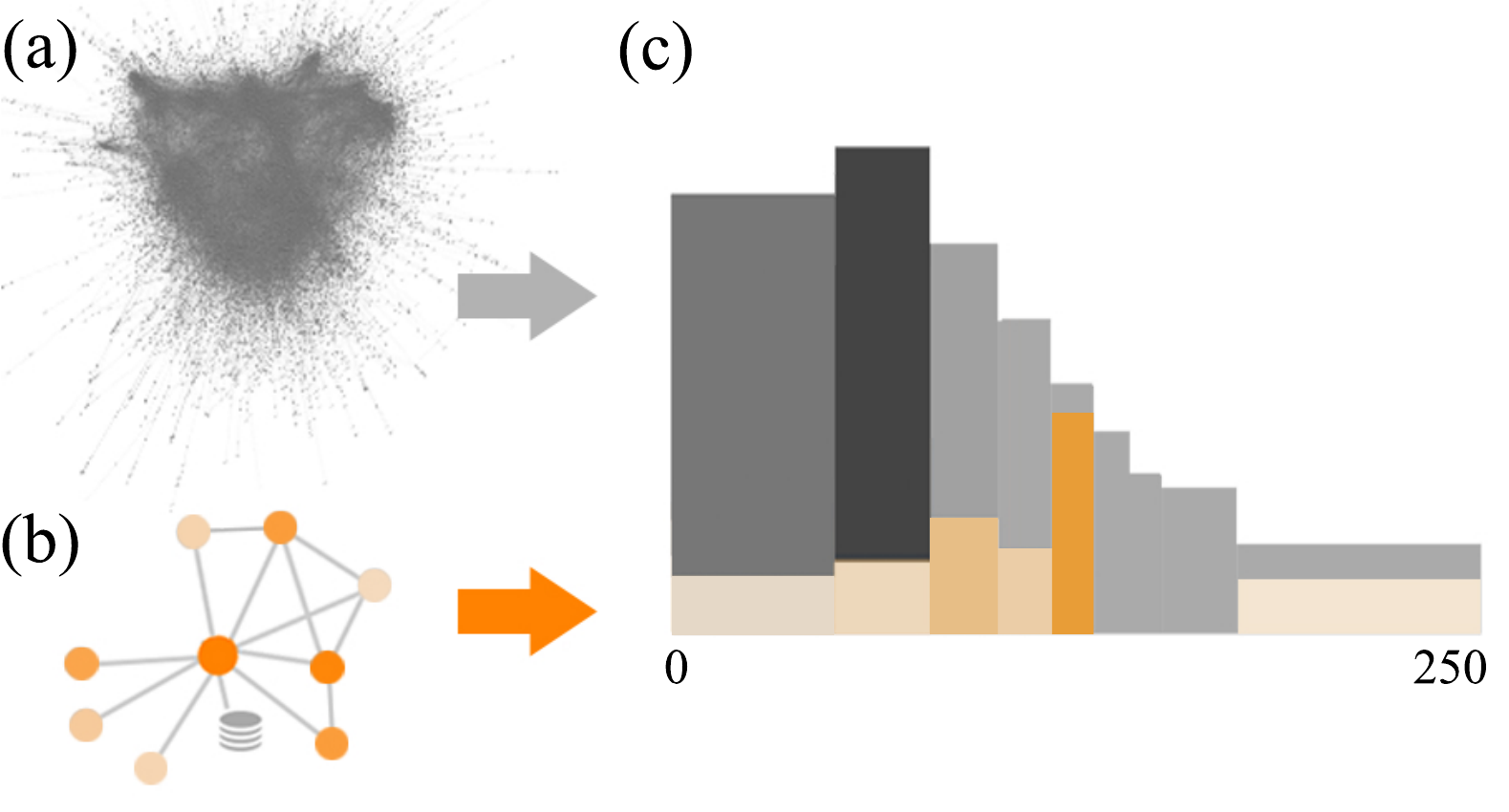}
\caption{Local and global distribution histograms are both essential to \name.
Local histograms (orange bars) are representations of feature distributions in a single node's egonet.
The global distributions (gray bars) depicts the corresponding feature's distribution across the whole graph.
The difference between those two distributions reflects if a node is "unusual" or surprising compared to the majority in the graph.
}
\vspace{-10 pt}
\end{figure}
Through our experiments we have chosen Jensen-Shannon (JS) divergence, a symmetrical version of Kullback-Leibler divergence to construct our surprisingness metric.
JS divergence works well, because the resulting output is in bits so the divergences of several features can be easily combined into a single score. 
We measure surprise by determining the divergence of feature distributions $L_{i,j}$ over a node's neighborhood $V_a$ (1 hop), from the global distributions of features $G$ (see Equation \ref{eqn:surprisescore}).
From these scores we select the top-$k$ most surprising nodes (Equation \ref{eqn:surprisek}).

Given the JS Divergence or information radius between two distributions $P$ and $G$: 
\begin{equation}
D_{\JS}(P || G) = \frac{1}{2}D(P||Q) +\frac{1}{2}D(G||Q),
\label{eqn:jsdiv}
\end{equation}
where $Q = \frac{1}{2}(P+G)$ and $ D(P||G)$ is the KL divergence for discrete distributions:
\begin{equation}
D(P||G) = \sum_b P(b) \log\frac{P(b)}{G(b)}
\label{eqn:kldiv}
\end{equation}
In Equation \ref{eqn:kldiv} we use base 2 so that $0\le D_{\JS}(P||G) \le 1$.
For a fresh node $v_a$, whose neighbors are not yet visualized we first compute the surprise-score,  $s_i$, of all neighboring nodes $v_i \in N(v_a)$:
\begin{equation}
s_i = \sum_{f_j \in A} \lambda_j D_{\JS}(L_{i,j}||G_j),
\label{eqn:surprisescore}
\end{equation}
where $L_j$ and $G_j$ are the local and global distributions of node-feature $f_j$ and $\lambda_j$ is a feature weight.
Weighted feature scores in Equation \ref{eqn:surprisescore} are used to lessen the impact of noisy features and to allow the user to lessen the contribution of a feature manually.
The $s_i$ scores are composed into $\hat S_a$, which holds all the scores for the neighbors of initial node $v_a$.
We can find the most surprising $k$-nodes by looking for the largest divergence from the global:
\begin{equation}
 \arg\max_{1 \dots k} \hat S_a 
\label{eqn:surprisek}
\end{equation}
This yields the top-$k$ most surprising nodes among the neighbors of node $v_a$.
Since both the local-neighborhood feature distributions and the global feature distributions are static, the surprise scores can be precomputed to improve real time performance.
We precompute and store surprise in \name to improve performance.

\subsection{Ranking by Subjective Interestingness \label{sec:interest}}
We track the user's behavior and record a user profile as they explore their data.
Each clicked node offers valuable details into the types of nodes in which the user is interested.
This forms distributions $U_j$ for each feature $f_j$.

To rank the user's interest in the undisplayed neighbors of node $v_a$ we follow a similar approach as Equation \ref{eqn:surprisescore}:
\begin{equation}
r_i = \sum_{f_j \in A} \lambda_j D_{\JS}(L_{i,j}||U_j) ,
\label{eqn:interestscore}
\end{equation}
where $U_j$ is the distribution of feature $f_j$ from the user's recent node browsing.
In this case we want the local distributions that match better the user's current profile; i.e. we want the smallest possible divergences:
 \begin{equation}
 \arg\min_{1 \dots k} \hat R_a
 \label{eqn:interestk}
\end{equation}
Since this suffers from the cold-start phenomenon, because a user will not have a profile until they have explored some nodes,
our remedy is to simply start the suggestions with surprising and important nodes, until the user has investigated several nodes.

\section{Our Integrated Approach}
\label{sec:interface}
The visual aspects of \name  with the algorithmic rankings described previously.
In this section we cover the graph visualization and design of our approach.

\begin{figure*}[t]
\centering
\includegraphics[width=\textwidth]{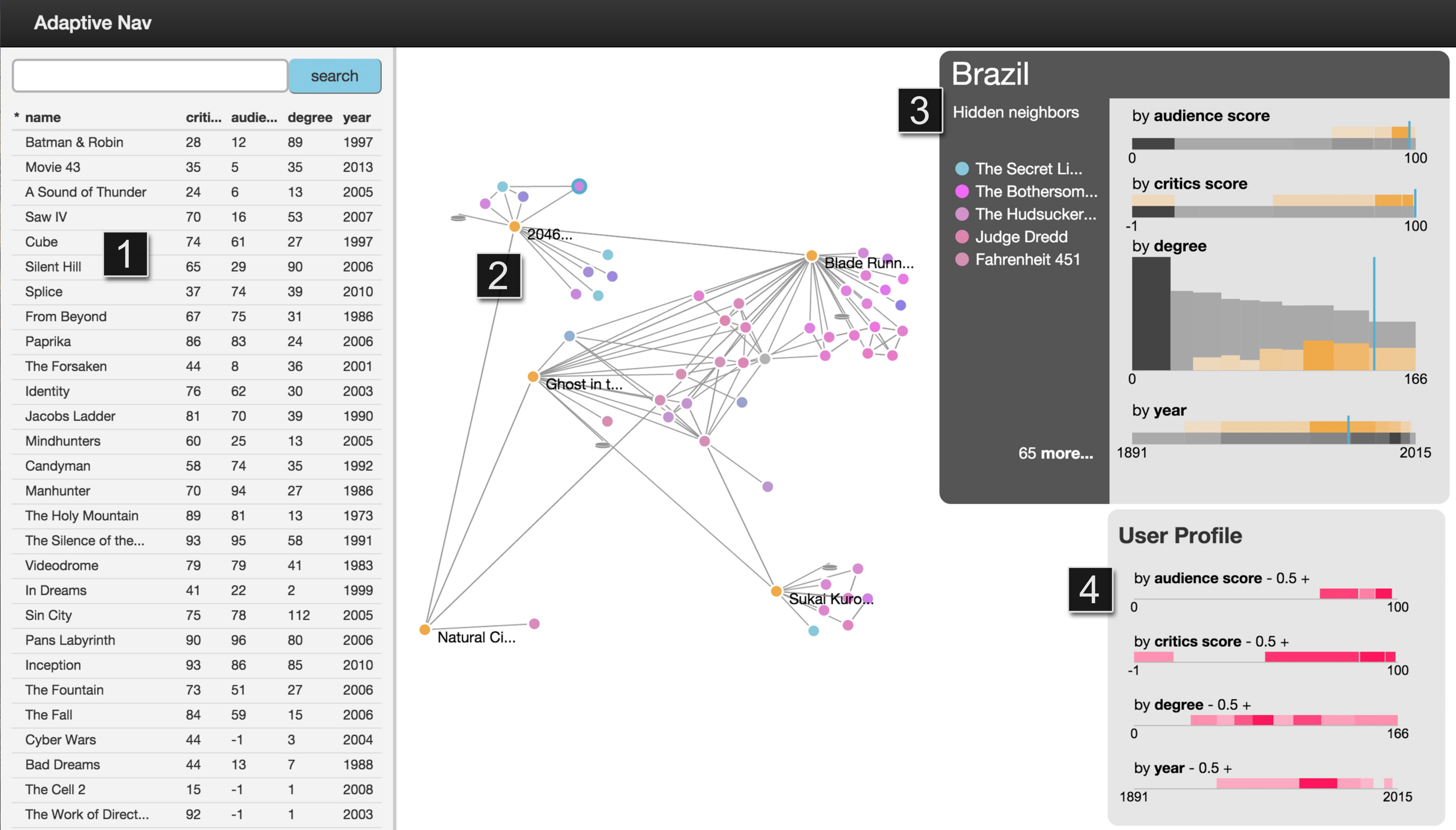}
\caption{The \name user interface displaying an explored portion of the RottenTomatoes similar-movie graph. The user has traversed several films (shown as orange nodes with titles) and \name has displayed a subset of the relevant neighbors (the nodes  with colors ranging from blue to red) and their connectivity.  1: The \textit{Table View} provides a conventional summarization of the already explored nodes and some of their features. 2: The \textit{Graph View} shows the connectivity of similar films as the user explores (it is linked with the table so that a selected node is highlighted in both views. 3: The \textit{Neighborhood Summary} shows the currently hidden nodes, by their feature distributions, for a selected film. 4: The \textit{User Profile} demonstrates a heat-map or flat histogram of the features the user has covered so far in their exploration. Figures are viewed best in color.}
\label{fig:ui_main}
\end{figure*}

\subsection{Components}
We have created an adaptive graph exploration tool, \name, for performing fast and intuitizve exploration of graph datasets.
\name was designed especially for graph tasks that require node-level details.

\name's user interface as shown in Figure \ref{fig:ui_main} has four key elements:
The first main area is the \textbf{Table View} (1),  a conventional table showing the currently displayed nodes and some of their features.  This provides sortable node-level information.
The central area is the \textbf{Graph View} (2). 
It is an interactive \textit{force-directed graph layout} that demonstrates the structure and relationships among nodes as the user explores. 
Coloring the nodes is used to encode the surprise and interest based on the user's current exploration. 
Visual scalability is an enormous challenge for nodes with many neighbors. 
Because we can only show a subset of the total nodes, we created the \textbf{Neighborhood Summary} (3).
The neighborhood summary allows a user to investigate the feature distributions of its currently undisplayed neighbors.
It presents the user with feature \textit{heat maps}\footnote{While histogram encodes value of each bin as height, heat map uses darkness to represent values with equal height. The main advantage of heat map over histogram is its compact representation, which helps us save space.} that summarize the distributions of hidden nodes.
When clicked, the heat maps turn into conventional distribution plots (histograms), where a user can compare the local neighborhood (orange) and the global (gray).
This lets a user quickly select new nodes based on their feature values and get a quick summary of this node's neighborhood. 
As a user explores, we construct and display a summary profile of the important features they have covered in the \textbf{User Profile} view (4).
The user profile view suggests high-level browsing behavior to the user; allows for better understanding of where the user-interest ranking comes from; and allows them to adjust if they want to ignore certain features in the interest ranking.


\subsection{Design Rationale}
In the following paragraphs, we discuss  \name's contributions to graph exploration.

\begin{figure}[bt]
\centering
\includegraphics[width=.5\textwidth]{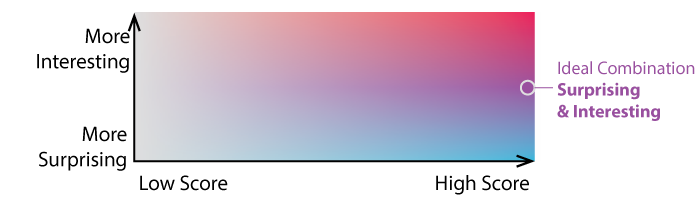}
\caption{\name's color encoding of user-interest and surprise using hue and saturation. The horizontal axis indicates the overall score of  a node; nodes with high scores will be located at the right of the chart and be colored with vibrant color. The vertical axis shows how much more user-interest score is than surprise score for a node.}
\label{fig:color_scale}
\end{figure}

\subsubsection*{Exploring and Navigating}
\begin{figure}
\centering
\includegraphics[width=.45\textwidth]{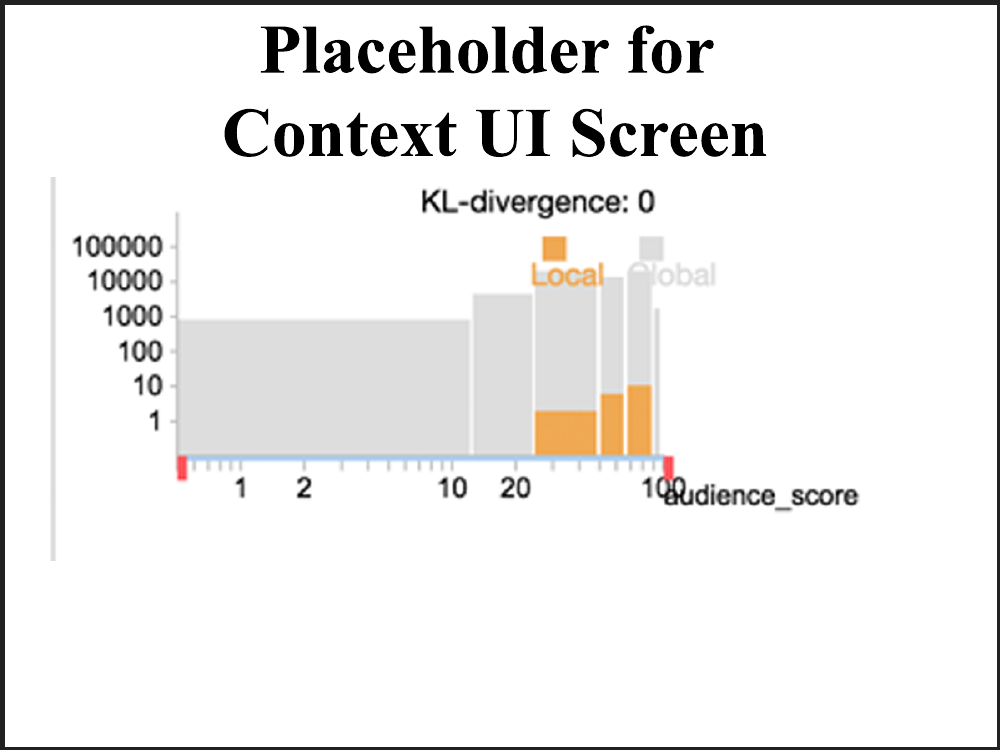}
\caption{\name's neighborhood summary view, which displays the top ranked neighbors of the given node (left) and distributions of the current neighborhood's features (right). Each feature is displayed by a compact heat map, which can be expanded into a histogram. Each heat map shows both the global distribution (gray) and the node's neighborhood's distribution (orange).}
\label{fig:ui_context}
\end{figure}

We use the term \textit{graph navigation} to refer to the act of traversing graph data with a known destination or objective.
\textit{Graph exploration} is similar to navigation except that the user has little or no particular destination.
One of our central design goals is to facilitate both exploration and navigation. 
We facilitate navigation through adaptation and exploration by filtering out unsurprising and unimportant nodes while still providing crucial feature details for hidden nodes via the Neighborhood Summary window.
As shown in figure \ref{fig:ui_context}, the user can have a summarized view of a mouse-hovered nodes where the top ranked hidden neighbors, local distribution and global distribution are displayed. This lets the user be able to go through many nodes quickly in seek of nodes with desired feature and add those desired nodes to graph view easily from the top hidden node list or bin.

\subsubsection*{Show the Best First}
Keeping the graph view from becoming a hairball means only showing relevant, surprising, and interesting nodes.
Importance, surprise, and user-interest are all important aspects of discovery, so we blend them into the results that are shown first to the user. Figure \ref{fig:color_scale} illustrates how we visually encode the interest-surprise difference by hue and the sum of both scores by saturation. Nodes ranked high tend to have brighter color closer to purple, which becomes a clear visual cue for the user to quickly identify desired nodes.
\name is almost completely free of parameters, making it simpler for users to explore their graphs.

\subsubsection*{Adaptive and Adjustable}
Because user-interest varies greatly across users and even time, our design must be able to track the user's exploration behavior in order to approximate what is motivating them.
Adapting as the user explores helps provide critical insight into users' latent objectives, because they can see how they have explored and also may find what they seek.
During exploration, the users profile updates dynamically to illustrate a summary of their feature traversal, while the graph view provides the topological traversal.
Even though there is no such necessarity to preset any parameters in order for our adaptive algorithm to work and the rankings are done in a black-box fashion during users' exploration, we allow them to directly manipulate the balance of features used in the interest calculation as well as the choice of which aspects form the ranking. Thereby when deciding to focus on only some of the features or weigh some feature less after having done some exploration with an established user profile, the user has the option to do so.


\subsection{The \ourmethodtitle Algorithm}

In this section, we summarize the process of finding top-$k$ most interesting and surprising neighbors in
Algorithm~\ref{alg:rankneighbors}.
Whenever a user selects a node to explore, we rank its neighbors based on surprise and subjective interestingness we explained in Sec. \ref{sec:surprise} and \ref{sec:interest}.
For each of the neighbors, we compute surprise and interest scores for each feature and aggregate them based on feature weights $\lambda_j$.
We blend those scores, and finally, nodes with $k$ highest scores will be returned.

\label{sec:facet_alg}
\begin{algorithm}[h]
\DontPrintSemicolon
\SetKwInOut{Input}{input}\SetKwInOut{Output}{output}

\Input{node $v_a$, precomputed histograms for: global feature distributions $G$ and user profile distributions $U$} 
\Output{$k$ most interesting and surprising neighbors of node $v_a$} 
\BlankLine 
\If{deg($v_a$) $\ge$ 1000}{
 $ V_a \leftarrow$ top 1000 neighbors of $v_a$ by highest degree \\
}  \Else {
$V_a \leftarrow$ all neighbors of $v_a$
}
  
\BlankLine
\ForAll{nodes $v_i$ in $V_a$}{
\ForAll{features $f_j$ in $A$}{
    $  s^{(j)}_i = D_{\JS}(L_{i,j}||G_j)$\tcp*[r]{see Eq.~\ref{eqn:surprisescore}} 
    $r^{(j)}_i =  D_{\JS}(L_{i,j}||U_j)$\tcp*[r]{see Eq.~\ref{eqn:interestscore}} 
    $t^{(j)}_i = w_s s^{(j)}_i + w_r (1-r^{(j)}_i )    $
 }
 $\hat T_a[i] =  \sum_{f_j \in A} \lambda_j t^{(j)}_i $
 }
  $T_{\Nodes} = \arg\max_{1\dots k} \hat T_a$ \\

\caption{RankNeighbors}
\label{alg:rankneighbors}
\end{algorithm}

\section{Experiments}
\label{sec:experiment}
In this section, we evaluate the effectiveness and speed of \name.
Our goal with \name is to facilitate user discovery by directing users towards regions of the graph with features suited to their interests.
Our goal is \textit{exploratory search} \cite{marchionini2006exploratory}; we are not trying to improve the accuracy of individual node recommendations, so it is not relevant to use canonical evaluation metrics like precision, recall, MAE, and RMSE~\cite{fouss2007rwrrec, wang2006graphrec, herlocker2004evaluating}. 
Instead, we demonstrate the effectiveness of our approach with (1) three case studies that investigate the results of our algorithm and (2) a comparison of our scoring with canonical node ranking techniques.

\subsection{Datasets}
We consider four large graphs to evaluate our method. First, we use the RottenTomatoes\footnote{\url{http://rottentomatoes.com/}} (RT) movie dataset, an attributed graph that contains basic information per movie (e.g., released year), as well as users' average ratings and critics scores. Second, we use the Google Web network, the DBLP co-authorship graph, and the YouTube network datasets from the SNAP repository\cite{leskovec2014datasets}.
We give the base characteristics of these graphs in Table \ref{table:datasets}.
\name ignores nodes with zero degree, because they aren't very useful for graph exploration.

\begin{table}
\centering
\small
\begin{tabular}{l rr}
\toprule
\textbf{Network} & \textbf{Nodes} & \textbf{Edges}  \\
\midrule
Rotten Tomatoes & 17\,074 & 72\,140\\
DBLP & 317\,080 & 1\,049\,866 \\
Google Web & 875\,713 & 5\,105\,039\\
Youtube & 1\,134\,890 & 2\,987\,624\\ \hline
\end{tabular}
\caption{Graph datasets used in our speed testing. They were picked for their variety in size and domain.  
}
\label{table:datasets}
\end{table}

\vspace{10 pt}
\subsection{Runtime Analysis}
Next, we evaluate the scalability of \name over several graphs. In particular, we will demonstrate its linear cost in the exploratory rankings. We note that the goal of this evaluation is mostly proof-of-concept -- in this paper it is not our focus to get the lowest possible ranking time, which could be achieved via better hardware and further engineering.

Guided graph exploration requires sub-second rankings in order to remain smooth and reactive to user input.
We must have rankings ready to display in under a second and ideally under half a second.
This is why we have chosen to treat nodes in the tail of the degree distribution separately than their modest degree neighbors. 

\begin{figure}
\centering
\includegraphics[width=.49\textwidth,trim={0.22in 0 0.22in 0},clip]{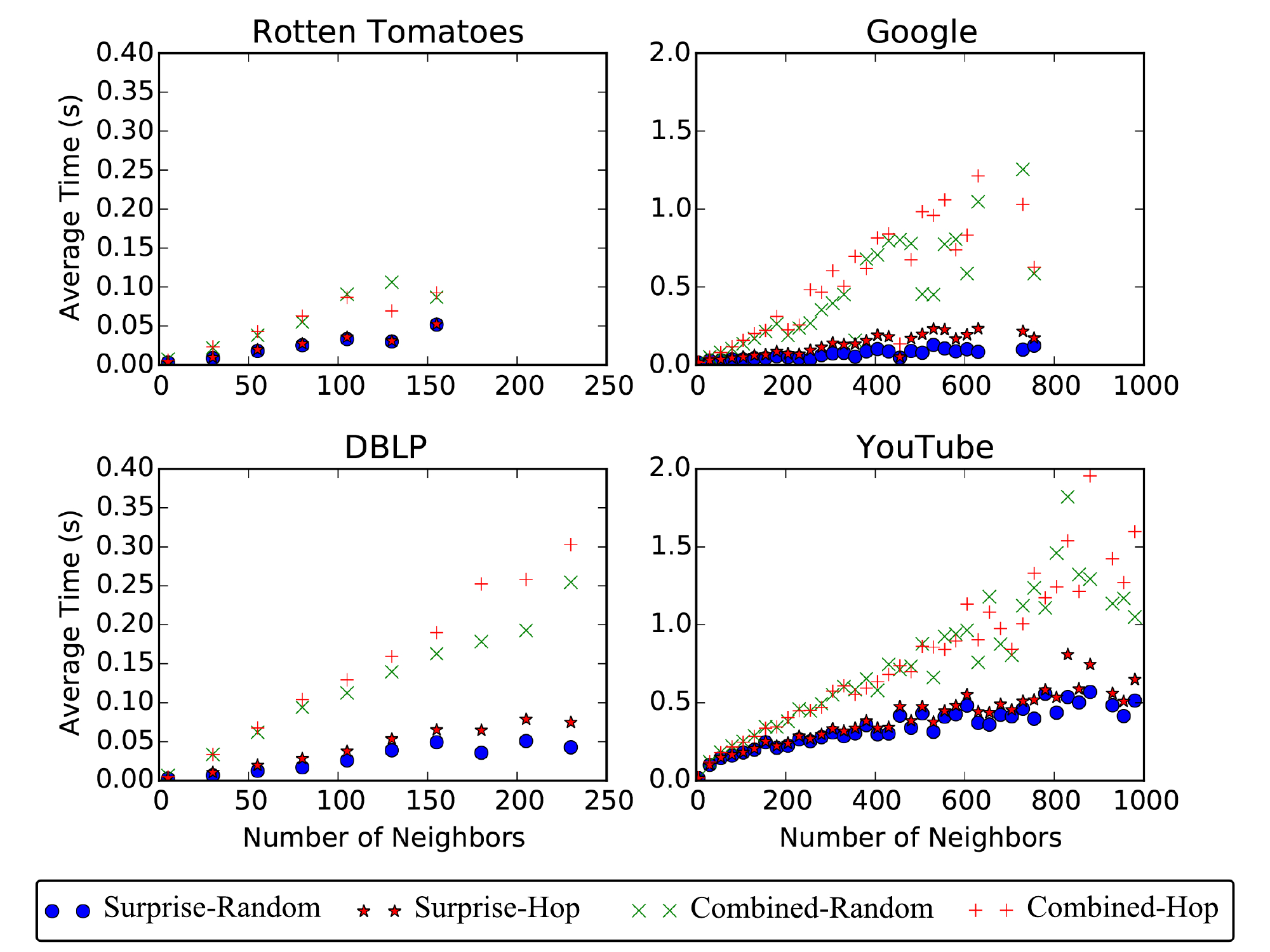}
\caption{\name ranks neighbors in linear time, which is necessary to handle the large numbers of nodes that a user may explore. 
We show the average time to calculate the JS divergence for surprise and the combination of surprise and interest over a neighborhood of size $n$. \name uses the combined ranking if the user has explored enough to have a user profile from which we can extract the subjective interest ranking.  
We tested with contiguous node ordering to simulate normal exploration and random ordering to simulate a user searching using the table view.}
\label{fig:timing}
\vspace{-10pt}
\end{figure}

We have analyzed the runtime of \name, in Figure \ref{fig:timing}, using the graphs from Table \ref{table:datasets}; all but the RT graph used eight synthetic features.
In our tests we use both random ordering and contiguous node ordering, displayed as Rand and Hop in Figure \ref{fig:timing}.
Random ordering simulates using the search functionality while the hop ordering simulates hopping from one node to its neighbors during exploration.
During the hop tests, the choice for the next hop must be among connected nodes that have not yet been traversed.  
High degree nodes have a  higher chance of being selected and account for the fact that hop sometimes is slower than random in Figure \ref{fig:timing}.
The graphs we tested demonstrate that the cost of the ranking is linear in the number of neighbors in the neighborhood.
Our ranking requires both a value lookup and a single JS divergence calculation for each node and for each feature.

\begin{figure}
\centering
\includegraphics[width=.5\textwidth]{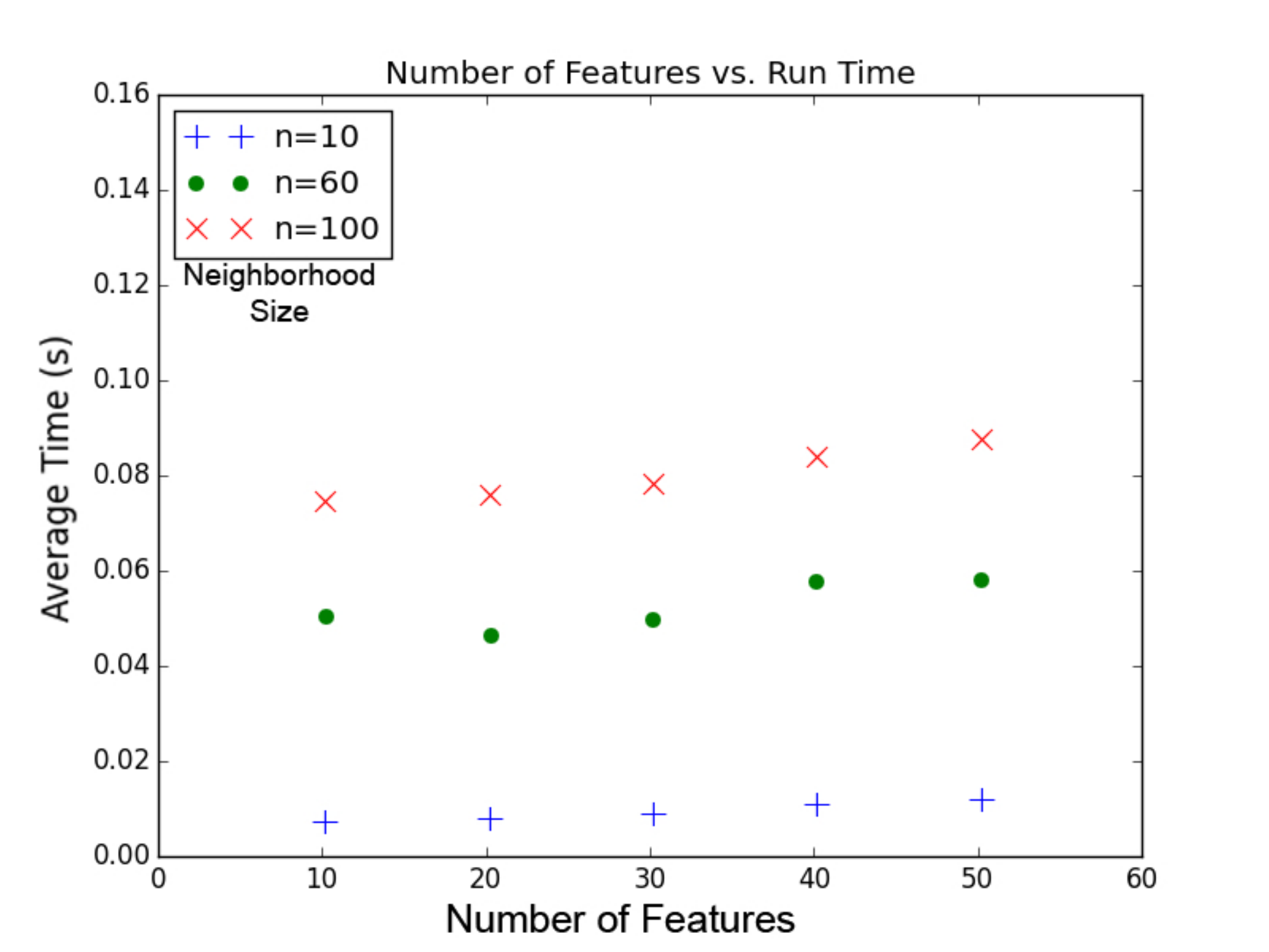}
\caption{\name scales linearly in the number of features, because the ranking requires exactly one JS divergence calculation per neighbor for interest and constant time lookup for surprise. Because of the compact structure of our MDL bins, the incremental cost of adding additional features is quite low.}
\label{fig:features}
\vspace{-10pt}
\end{figure}

As mentioned in Section \ref{sec:model}, the surprise scores are precomputed and can be accessed quickly. 
The cost to rank neighbors comes largely from the interest scores which cannot be precomputed.
The cost, in JS divergence calculations, is $O(n\cdot f)$, and  is asymptotically linear in both the number of neighbors $n$ and the number of features $f$.
Given our use of MDL histograms for the features, we are able to scale the number of features at low linear incremental cost (see Figure \ref{fig:features}).
Each neighbor only requires exactly one JS divergence calculation per feature (comparing the user profile and the local distribution).

Since many graphs contain triangles, it is very likely that redundant calls will be made during a user's exploration. 
We use this to our advantage and \textit{cache} the distributions for each visited node rather than refetching them each time.
Graphs with higher clustering coefficient may achieve better caching performance.
For all but the Youtube graph, the caching became memoizing as the entirety of the nodes could fit in the cache.

\begin{table*}[t]
\small
\centering
\begin{tabular}{l lrrrr lrrrr} 
\toprule
 & \multicolumn{5}{c}{\textbf{Example I}}&\multicolumn{5}{c}{\textbf{Example II}}   \\[0.5em]
\textbf{Current Movies} & \multicolumn{5}{c}{\textbf{Blade Runner}}&  \multicolumn{5}{c}{\textbf{Toy Story}}\\ 
\cmidrule(lr{1pt}){1-1}
\cmidrule(lr{4pt}){2-6}
\cmidrule(l{4pt}r){7-11}
  \multirow{2}{*}{\textbf{Movies Visited}} 
& \multicolumn{5}{c}{Waterworld, Braveheart, Pulp Fiction,}
& \multicolumn{5}{c}{A Bug's Life, Kung Pu Panda, Jumanji,} \\
& \multicolumn{5}{c}{The Crow, Fargo}
& \multicolumn{5}{c}{The Incredibles, How to Train Your Dragon} \\
\cmidrule(lr{1pt}){1-1}
\cmidrule(lr{4pt}){2-6}
\cmidrule(l{4pt}r){7-11}
\textbf{Top-$k$ by}&  \textbf{Title} & \textbf{I} & \textbf{PR} & \textbf{ BC} & \textbf{ EV} & \textbf{Title} & \textbf{I} & \textbf{PR} & \textbf{ BC} & \textbf{ EV}\\ 
\cmidrule(lr{1pt}){1-1}
\cmidrule(lr{4pt}){2-6}
\cmidrule(l{4pt}r){7-11}
  \multirow{5}{*}{\textbf{Interest}}
& L.A. Confidential & 1.06 &2.23 & 
264k 
&7.10 & Monsters University& 0.54 & 0.76 & 1070&29.3 \\
& Sin City & 1.15 &4.34	& 833k 
&44.7&   Rio &0.63 		&1.66&25464&57.4  \\
&  Dredd & 1.20 &1.74		&55k 
& 43.3& Toy Story II& 0.70 	&1.84&55486&59.1  \\
& V for Vendetta & 1.23&2.99&620k 
& 43.5 & The Iron Giant&  0.71 	&1.43&54079&46.3 \\
& Heat & 1.23 &4.18		&385k 
&11.0& ParaNorman &  0.71 	& 0.94 & 23291&15.1 \\ 
\cmidrule(lr{1pt}){1-1}
\cmidrule(lr{4pt}){2-6}
\cmidrule(l{4pt}r){7-11}
\textbf{Top-$k$ by}&  \textbf{Title} & \textbf{S} & \textbf{PR} & \textbf{ BC} & \textbf{ EV} & \textbf{Title} & \textbf{S} & \textbf{PR} & \textbf{ BC} & \textbf{ EV}\\ 
\cmidrule(lr{1pt}){1-1}
\cmidrule(lr{4pt}){2-6}
\cmidrule(l{4pt}r){7-11}
  \multirow{5}{*}{\textbf{Surprise}}
&  The Creation of the Humanoids &2.93& 0.19 & 851 & 0.58 & Buzz Lightyear of Star Command & 3.15 & 0.14 & 12&2.14 \\
&  Demon Seed &2.75 	& 0.50 & 650
&1.54& Small Fry & 3.01 & 0.17 & 132&3.10 \\
& Natural City & 2.69 	&0.28 & 437	&0.74& Monsters University& 2.64 &0.76  &1070&29.3  \\
& Virtuosity & 2.64 		&0.35 & 668	&1.61 & Cloudy With a Chance of Meatballs &  2.46 &1.28 & 1027&42.5  \\
&  Soylent Green & 2.59 	& 0.57 & 6578 
&1.50&ParaNorman & 2.45 & 0.94 & 23291 & 15.1 \\ \bottomrule
\end{tabular}
\caption{Comparing \name's surprise and interest ranking with common importance rankings (I\&S is interest or surprise, PR is PageRank ($\times10^{-4}$), BC is betweenness centrality, EV is eigenvector centrality ($\times10^{-3}$). Each example has a selected film, a user profile at the time of selection, and the interesting and surprising neighbors.  }
\label{table:casestudy}
\end{table*}

\begin{figure*}[ht]
\centering
\subfigure[]{
   \includegraphics[width=.47\textwidth]{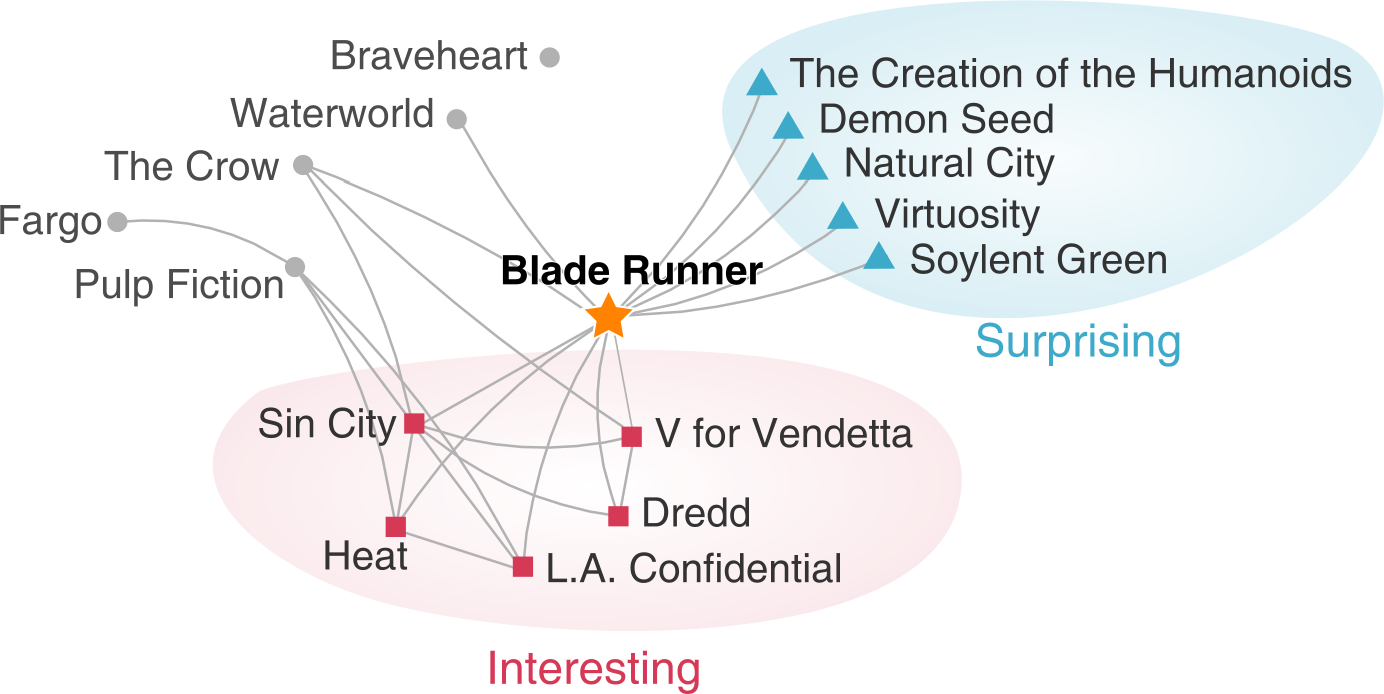}
   \label{fig:bladerunner}
 }
\qquad
 \subfigure[]{
  \includegraphics[width=.47\textwidth]{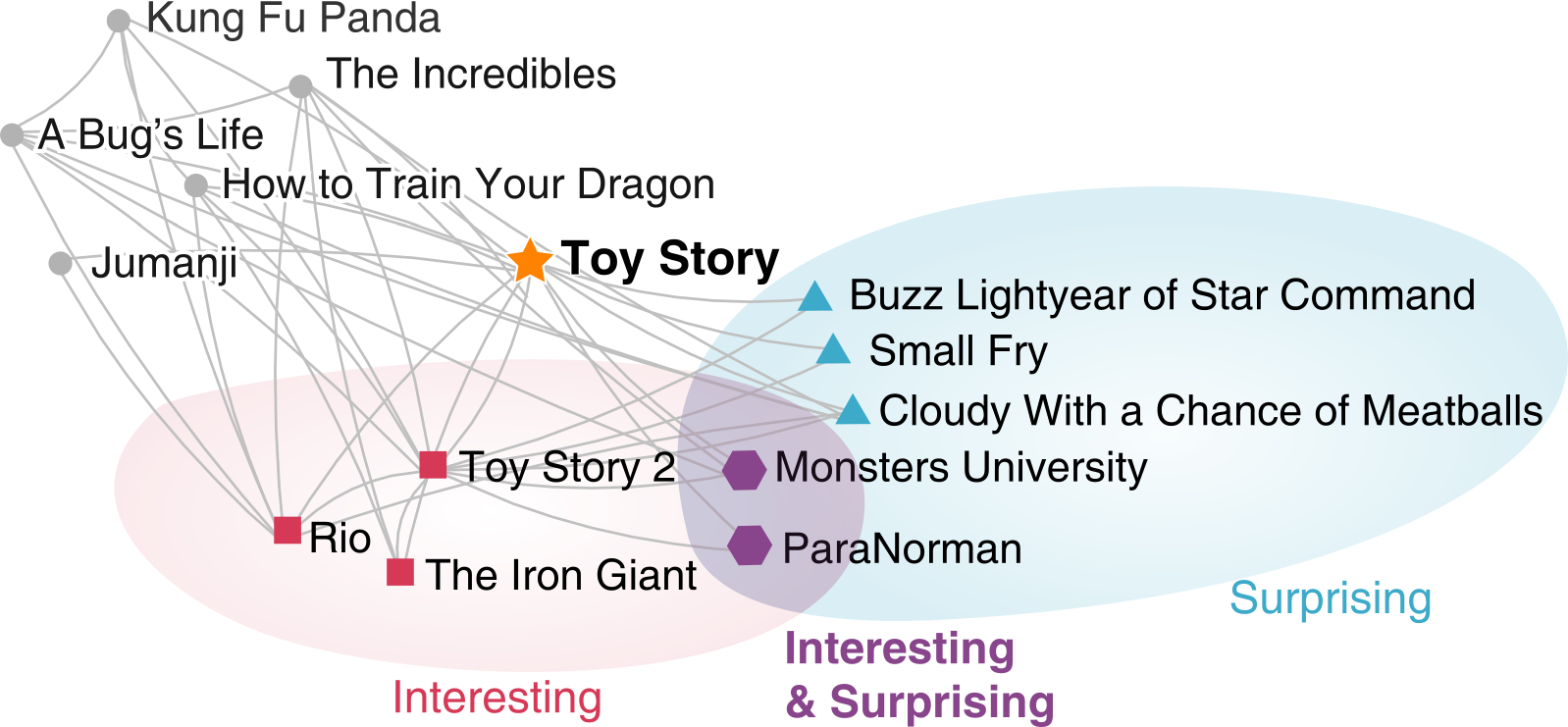}
   \label{fig:toystory}
}
\vspace{-10pt}
\caption{Visualizations of the (a) \textbf{Blade Runner} and (b) \textbf{Toy Story} case studies. The gray, circular nodes form a profile of films investigated by the user via graph exploration or text-search. The star nodes are the last-clicked node for which \ourmethod has produced the 5 most interesting (red), and surprising (blue) nodes. Nodes can be both surprising and interesting (purple).}
\label{fig:casestudygraphs}
\end{figure*}

\subsection{Case Studies}
Here, we describe three case studies using two graph datasets --- the first two using the Rotten Tomatoes (RT) movie graph, and the last one using the DBLP co-authorship graph.
These case studies illustrate how \ourmethod helps the user explore and graphs, incrementally gain understanding, discovers new insights, and visualize them.
\ourmethod adapts to the user to provide surprising and interesting rankings to help the user explore relevant parts of the graphs.


\subsubsection*{Movie Example I: Blade Runner}
Our first user John likes mid-90's post-apocalyptic and action films (shown as gray circles in Figure \ref{fig:bladerunner}, e.g., \textit{The Crow}, \textit{Waterworld}), many of which were well received by critics and audience alike.
John can add more films that he likes by exploring node-to-node or by the search utility.
After exploring a few films in such genre, John is particularly interested in \textbf{Blade Runner}.
Based on why John has explored, \ourmethod returns top-ranking surprising and interesting films.
The top-5 of each kind are displayed in Table \ref{table:casestudy} on the left and in Figure \ref{fig:bladerunner}.
These movies' interest (I) and surprisingness (S) scores are determined based on conventional measures of node importance; PageRank (PR), betweenness centrality (BC), and eigenvector centrality (EV).

Many of the surprising movies would not considered important by canonical approaches, partly because \ourmethod's surprise rank operates on both features and local graph structure rather than global structure.
Movies that are very heavily connected also face a higher chance of matching the global distribution and therefore being less surprising.

While the notion of surprise is often considered ``unimportant'' by conventional metrics, 
interest exhibits more variety in importance than surprise.
This is especially apparent in both this and the next case studies.
Here we have selected types of films that have really large viewership and represent a very large genre in modern film with many potentially similar movies.
Both the structure and features used in our subjective interest have lead the ranking towards more conventionally important nodes (consider the high PR, BC, and EV scores).

As the highest possible JS-divergence for a single feature is 1; the maximum divergence is the sum of feature weights $\Lambda$. 
In this case study, we used five features, so interest scores $\le 1.0$ suggest that the user profile is in relatively good agreement with the proposed node.
Nodes with very low interest divergence are strong candidates that their neighborhood will be of interest to the user.

\subsubsection*{Movie Example II: Toy Story}
Our second user Bonnie investigates several children's films that were criticized by the critics, but still enjoyed by audiences (all have a lower critics' score than audience score). 
Bonnie hops from node-to-node across the listed films in the order they appear in Table \ref{table:casestudy}.
She then selects \textbf{Toy Story}.
The results are displayed in Table \ref{table:casestudy} on the right and in Figure \ref{fig:toystory}.

As in the previous example, the surprising nodes tend not to be conventionally important.
Yet, consider the difference in importance of the interesting films versus the previous example, many of these suggestions have significantly lower importance.
The second profile has less coherent features and they do not draw the the interest ranking towards conventionally important nodes, despite that \textit{Toy Story} is a very well connected node.
Also note that \textbf{Monsters University} and \textbf{ParaNorman} are featured in both the top-5 surprising and interesting movies.
The surprise score and interest score are not counter to each other. 
In fact, this is an excellent example, because these films are both surprising and subjectively interesting!

\subsubsection*{DBLP Example III}
Our third example uses data extracted from DBLP, a computer science bibliography website \footnote{http://dblp.uni-trier.de}. 
The graph is an undirected, unweighted graph describing academic co-authorship. 
Nodes are authors. 
An edge connects two authors who have co-authored at least one paper.

\begin{figure}[]
\begin{center}
\begin{tabular}{c}
\includegraphics[width=0.45\textwidth]{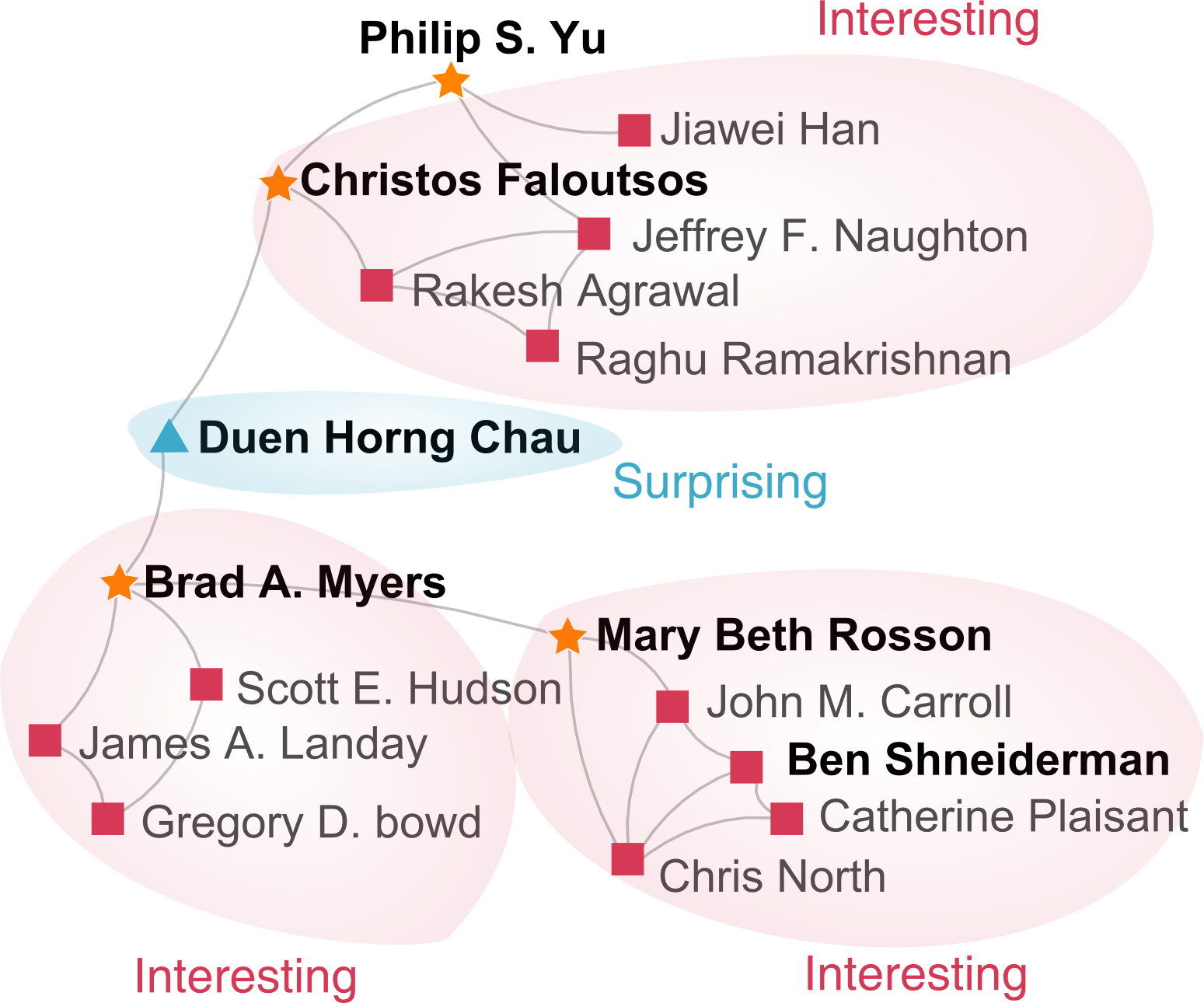}
\end{tabular}
\end{center}
\caption{
Visualization of our user Jane's exploration of the DBLP co-authorship graph. 
Jane starts with Philip Yu.
\ourmethod then suggests Christos Faloutsos and several others as prolific data mining researchers.
Through Christos, \ourmethod suggests Duen Horng Chau as a surprising author as he has published with both data mining and human-computer interaction (HCI) researchers, like Brad Myers.
Through Brad, \ourmethod helps Jane discover communities of HCI researchers, including Ben Shneiderman, the visualization guru.
\label{fig:dblp}
}
\end{figure}

Our user Jane is a first-year graduate student new to data mining research. 
She just started reading seminal articles written by Philip Yu (topmost orange star in Figure~\ref{fig:dblp}).
\ourmethod quickly helps Jane identify other prolific authors in the data mining and database communities, like \textit{Jiawei Han}, \textit{Rakesh Agrawal}, \textit{Raghu Ramakrishnan}, and \textit{Christos Faloutsos}; these authors have similar feature distributions as Philip Yu (e.g., very high degree).
Jane chooses to further explore \textit{Christos Faloutsos}'s co-authors.
\ourmethod suggests \textit{Duen Horng Chau} as one of the surprising co-authors, who seems to have relatively low degree (i.e., few publications) but has published with highly-prolific co-authors.
Among these is \textit{Brad Myers} (leftmost orange start in Figure~\ref{fig:dblp}), who publishes not in data mining, but in human-computer interaction (HCI).
This exploration introduces Jane to a new field, and she wants to learn more.
Using \ourmethod's interest-based suggestion, she discovers a community of co-authors who have published with Brad; 
among them, \textit{Mary Beth Rosson} further leads to another community of HCI researchers, which includes \textit{Ben Shneiderman}, the visualization guru!

\section{Related Work}
\label{sec:related}

Graph traversal and exploration has been investigated in several different areas related to networks.

\subsubsection*{Trails and Paths Through Networks}
The information retrieval community has proven an asset in analyzing the web-browsing paths users covered as they explored the web.
\textit{Click trail analysis} has been used to analyze the website-to-website paths of millions of users in order to improve the ranking of search results \cite{bilenko2008trail, singla2010trail}.
 
Intermediate sites and destinations of common trails can be used successfully as search results themselves as in \cite{white2010searchresults}.
If a user is following a common trajectory they can even be ``teleported'' to the destination page \cite{teevan2004directedsearch}.
However, in our case we do not have millions of explored paths through our input network and cannot directly rely on the aggregate analysis of trails used above.

West et al. analyzed Wikipedia users' abilities and common patterns as they explored Wikipedia \cite{west2012human}.
They observed that users would balance between a conceptually simple solution at the cost of efficiency--there were more direct routes to their target article that made less sense than they path they chose.
Their system also used trail analysis in order to try to predict where a user would go based on the user's article-trail features.

\subsubsection*{Degree of Interest}
The visualization community has also investigated local graph exploration.
Bottom-up exploration first appeared in \cite{furnas86fisheye}, a tool for exploring hierarchies using a ``degree of interest'' (DOI) function to rank the relevance of currently undisplayed nodes.
The idea of DOI was later expanded by \cite{van2009search} to apply to a greater set of graph features.
The Apolo system \cite{chau2011apolo} further improves on it to  allow  users to freely define their own arbitrary number of clusters, which it uses to determine what to show next, through the Belief Propagation algorithm.
Recently, the DOI idea is applied to time-varying settings, to capture salient temporal graph changes~\cite{abello2014modular}.

We have built on the idea of using a DOI to determine the ranking for what nodes we show users; however, we use a \textit{dynamic DOI}  function which changes to suit the browsing behavior of the users as they explore their data.

\subsubsection*{Surprise and Serendipity}
Algorithms like Oddball \cite{akoglu2010oddball}, an unsupervised approach to detect anomalies in weighted graphs, can be used to detect surprising nodes. 
Akoglu et al.~focused on anomalies based on edge weighting and not on node-level features.
The TANGENT algorithm by Onuma et al. is a parameter free technique used to discover surprising recommendations \cite{onuma2009tangent}.
They measure the surprise in their model by measuring the amount of horizon-broadening each new node imparts, where the horizons are edges going to different clusters of users and films.


Similar to our approach to comparing local and global distributions, Parameswaran et al. \cite{parameswaran2013seedb} presented the idea of finding interesting dimensions in the context of data cube by comparing their distributions to those aggregated.

Andre et al. \cite{andre2009x} studied the potential for using serendipity in Web search and the effects of personalization on that potential. 
They found out that serendipity could be useful for certain queries, and personalization, ideas which we leverage in \name. 


De Bie~\cite{debie:11:inftheoryframework,debie:11:dami} proposed a general framework for measuring the interestingness of data mining results as their log-likelihood given a Maximum Entropy (MaxEnt) distribution 
based on the background knowledge of the user. Instead, we consider the Jensen-Shannon divergence between the local and global histograms as the surprisingness of a node. Our notion of subjective interestingness comes closer to that of Tatti and Vreeken~\cite{tatti:12:apples}, whom proposed to iteratively update the model, to avoid redundancy. Our goal is orthogonal, as we  are specifically interested in identifying nodes that are similar to those the user chose to explore.

\section{Conclusion}
\label{sec:conclude}
In this work, we presented \name, an integrated approach that combines visualization and computational techniques to help the user performs adaptive exploration of large graphs. 
\name overcomes many issues commonly encountered when visualizing large graphs.
\name shows the user only the most subjectively interesting material as they explore. 
We do this by ranking the neighbors of each node by surprisingness and subjective interest based on what the user has explored so far.
The surprisingness rankings are determined by the divergence of local feature distributions from the background feature distributions.
The subjective interest rankings arise from the divergence between the local feature distributions and the user's current profile, which represents their exploration, in node-features, up to that point.

 Our \name algorithm is scalable and is linear in the number of neighbors and linear in the number of features.
 We demonstrated the effectiveness of \name through case studies using the RottenTomatoes movie graph and the DBLP co-authorship graph, and comparison with canonical importance ranking measures.
 The user interest ranking influences the direction of a user's exploration by showing the best matches to their current taste in nodes.
 Despite the old adage, you can see the graph through its nodes.

\section*{Acknowledgments} 
This material is based upon work supported by the National Science Foundation under Grant No. IIS-1217559 and National Science Foundation Graduate Research Fellowship Program under Grant No. DGE-1148903.
Funding was provided by the U.S. Army Research Office (ARO) and Defense Advanced Research Projects Agency (DARPA) under Contract Number W911NF-11-C-0088. 
Jilles Vreeken is supported by the Cluster of Excellence ``Multimodal Computing
and Interaction'' within the Excellence Initiative of the German Federal
Government. 
James Abello acknowledges support from mgvis.com, DIMACS, and NSF grant CCF-1445755.

%
\balance
\bibliographystyle{abbrv}
\bibliography{bib/abbrev,bib/paper,bib/bib-jilles}  
%
%

\end{document}